# Broadband enhancement of spontaneous emission from nitrogen-vacancy centers in nanodiamonds by hyperbolic metamaterials

M. Y. Shalaginov[1,2], S. Ishii[1,2,3], J. Liu[1,2], J. Liu[4], J. Irudayaraj[4], A. Lagutchev[2], A. V. Kildishev[1,2], and V. M. Shalaev[1,2,a]

[1]School of Electrical and Computer Engineering, Purdue University, West Lafayette, IN 47907, USA

[2]Birck Nanotechnology Center, Purdue University, West Lafayette, IN 47907, USA

[3]Advanced ICT Research Institute, National Institute of Information and Communications Technology, Kobe, Hyogo, 651-2401, Japan

[4]Agricultural and Biological Engineering, Purdue University, West Lafayette, IN 47907, USA

We experimentally demonstrate a broadband enhancement of emission from nitrogen-vacancy centers in nanodiamonds. The enhancement is achieved by using a multilayer metamaterial with hyperbolic dispersion. The metamaterial is fabricated as a stack of alternating gold and alumina layers. Our approach paves the way towards the construction of efficient single-photon sources as planar on-chip devices.

[a] Electronic mail: shalaev@purdue.edu

Nitrogen-vacancy (NV) centers in diamond have recently attracted widespread attention from the quantum optics community.[1] It has been shown that NV centers in diamond generate stable, broadband, and anti-bunched emission at room temperature, which makes it possible to use them as an efficient source of single-photons.[2] Also importantly, an NV center can serve as a key component in a quantum bit (qubit) for future information technology. Such centers are able to store quantum information for a significant amount of time, and be read out optically.[3] The efficiency of the NV center both as a single-photon generator and an element of a qubit is directly related to its spontaneous emission rate. Other important applications of NV centers, such as nanoscale electric[4] and magnetic[5,6] field sensors, will also strongly benefit from higher emission efficiency and an increased flux of single photons.

According to Fermi's golden rule, the radiative decay rate depends on both the internal properties of an emitter and the density of electromagnetic modes in the environment. Therefore, engineering the light source environment (Purcell effect) can lead to improvement of the spontaneous emission rate.[7] So far, this idea has been implemented with resonant structures, such as optical resonators,[8,9] photonic crystal cavities,[10] and plasmonic apertures,[11] which are all bandwidth-limited. In this work, we experimentally study non-resonant broadband enhancement of the emission from NV centers coupled with hyperbolic metamaterials (HMMs).[12-14]

Our approach takes advantage of the large photonic density of states (PDOS) in a broad range of wavelengths, which is a striking property of HMMs.[15,16] The PDOS, similar to its electronic counterpart, can be quantified as the volume in k-space between isofrequency surfaces $\omega(k)$ and $\omega(k)+d\omega$. For extraordinary waves in a uniaxial anisotropic medium with dielectric tensor $\varepsilon = \text{diag}[\varepsilon_{\|}, \varepsilon_{\|}, \varepsilon_{\perp}]$, iso-frequency surfaces are defined by the following dispersion relation:

$$\omega^2 / c^2 = k_{\|}^2 / \varepsilon_{\perp} + k_{\perp}^2 / \varepsilon_{\|}, \qquad (1)$$

where subscripts "$\perp$" and "$\|$" indicate the directions perpendicular and parallel to the plane of anisotropy, respectively. In case of conventional materials with $\varepsilon_{\perp}, \varepsilon_{\|} > 0$, PDOS is equivalent to an infinitesimally thin ellipsoidal shell in k-space. However, in a medium with hyperbolic dispersion $\varepsilon_{\perp}$ and $\varepsilon_{\|}$ are of the opposite sign which produces a hyperboloid shell whose volume is infinitely large in the effective medium limit (i.e. broadband singularity in PDOS appears). As a result, such a medium allows the propagation of spatial modes with an arbitrarily large wave vectors. However, in real structures, the maximum value of the wave vector is restricted by the size of the metamaterial unit cell (which is much smaller than the wavelength in a metamaterial).

The HMM could be achieved as a lamellar structure consisting of alternating subwavelength-thick layers of metal and dielectric[17,18] or as an array of nanowires embedded into a dielectric host matrix.[19,20] In this work, we fabricated multilayer HMMs consisting of 16 alternating layers of gold (Au) and alumina ($Al_2O_3$) with each being 19 nm thick, to form an overall thickness of 304 nm. Films of Au and $Al_2O_3$ were subsequently deposited on a 0.7-mm-thick glass substrate using electron-beam evaporation, with the top layer being Au. The thicknesses of the films were measured by quartz crystal microbalance. Dielectric functions of the fabricated HMM were retrieved by using spectroscopic ellipsometry (V-VASE, J.A. Woollam Co.).[21] The effective medium parameters are shown in FIG. 1. Note that $\text{Re } \varepsilon_{\perp} > 0$ and $\text{Re } \varepsilon_{\|} < 0$ is the signature of hyperbolic dispersion.

Based on our previous characterization,[22] we used nanodiamonds containing NV centers which were obtained in an aqueous suspension of 0.05% w/v from the Institute of Atomic and Molecular Sciences (Academia Sinica, Taiwan).[23] The average size of the crystals is reported to be 35 nm. According to manufacturer's information, the nanodiamonds were originally purchased from Microdiamant (Switzerland), and were additionally bombarded with 40-keV $He^{2+}$ ions and afterwards annealed at 800 °C for 2 hours to increase the concentration of NV centers. On average, there are ten centers per crystal.[24] Subsequently, a 100-times diluted aqueous suspension containing nanodiamonds was spin-coated at 3000 rpm onto a substrate. The dilution is necessary for increasing the distance between nanodiamonds and avoiding the formation of agglomerates when deposited on a substrate.

The emission spectra of NV centers was measured when placed in the vicinity of the HMM [Fig. 2]. In order to avoid photoluminescence quenching by a metallic surface, we spin-coated a 30-nm-thick spacer layer composed of diluted SU-8 photoresist (1:5 mixture with SU-8 thinner) at 5000 rpm on top of the uppermost gold layer followed by the dispersion of nanodiamonds. The fluorescence signal from NV centers was obtained using a custom-made scanning confocal microscope.[25] The 488-nm line of an Ar/Kr continuous-wave laser (0.5 mW) was used for excitation. The emitted signal was collected by 60x water immersion objective (UPlanApo/IR, Olympus with a NA of 1.2) and spectrally resolved using a Shamrock SR-303i-A spectrophotometer from Andor Technology equipped with a 300 grooves/mm, 500-nm blazed angle grating and a thermoelectrically cooled CCD camera (Newton DU920N-BR-DD). There was a 500-nm longpass filter in the signal path to block the excitation light and a pinhole to create a confocal configuration to reduce the background light. Results of the measurements [FIG. 2 (b)] demonstrate the broadband emission from NV centers when the nanodiamonds are placed on the surface of the HMM. Characteristic peaks are observed at the wavelengths of 575 nm and 637 nm corresponding to zero-phonon lines of $NV^0$ and $NV^-$ centers, respectively.[1]

In order to characterize the enhancement of emission due to the HMM, the ratio of the observed radiative decay rate ($\Gamma$) for the NV centers on the HMM and a reference sample was evaluated using the following method:[26]

$$\frac{\Gamma}{\Gamma_{ref}} = \frac{Q}{Q_{ref}} \frac{\tau_{ref}}{\tau} = \frac{I}{I_{ref}} \frac{A_{ref}}{A} \frac{\tau_{ref}}{\tau}, \tag{2}$$

where $\tau$ – excited-state lifetime, $Q$ – apparent quantum yield, $I$ – integrated photoluminescence intensity, and $A$ – absorption by an emitter. The reference sample consisted of nanodiamonds on a glass substrate, covered with a spacer layer. The values obtained for this sample are indicated with a subscript 'ref' in Eq. (2).

Lifetime measurements were carried out by using Micro Time 200 (PicoQuant GmbH) time-resolved scanning confocal microscope. A pulsed diode laser with a pulse duration of 50 ps, a wavelength of 465 nm, and a repetition rate of 40 MHz was used to excite the system. A laser beam carrying 2 mW of power was focused onto the sample using a 50x objective lens (NA 0.75). The fluorescence light emitted was collected with the same objective lens. The excitation light was blocked by a dichroic mirror and a band-pass filter (650-720 nm), which covers the majority of the NV center emission spectrum, to reject any remaining scattered laser light. The fluorescence light passed through a small aperture to enable confocal detection performed with SPCM-AQR avalanche photodiode from Perkin Elmer. The data acquisition was achieved by a time-correlated single-photon counting method which is based on the precise measurement of the time difference between the moment of excitation and the arrival of the first emitted photon at the detector. Data for the lifetimes is arranged in histograms shown in FIG. 3.

In the case of nanodiamonds on a glass substrate, the lifetime distribution is roughly symmetric with a mean value $\bar{\tau}$ = 20.89 ns and standard deviation $\Delta\tau$ = 1.15 ns [FIG. 3a]. In contrast, nanodiamonds on the HMM [FIG. 3b] produce non-symmetric distributions with $\bar{\tau}$ = 1.55 ns and $\Delta\tau$ = 0.95, reducing the average lifetime by a factor of 13.48. We also measured the lifetime when nanodiamonds are placed on a 300-nm-thick gold film and observed that the average lifetime is reduced by 6.3. The significant variation of lifetimes is expected to be due to the different orientations of NV center dipoles with respect to the substrate surface.[27]

Integrated photoluminescence intensity is assumed to be proportional to the total number of detected photons, i.e. to the histogram integral. Therefore, the ratio of integrated intensities in Eq. (2) was evaluated by the ratio of the corresponding histogram integrals [see TABLE I].

To obtain the absorption of the 465 nm excitation beam by the NV centers in nanodiamonds placed on glass and HMM substrates, we fabricated identical pairs of the substrates. Then, the original aqueous suspension of dispersed nanodiamonds (0.05% w/v) mixed equally with isopropyl alcohol (IPA) was spin-coated four times onto the one of a pair of substrates. IPA is required to make the spacer layer more wettable. Absorption (A) of the samples was obtained from the formula

$$A = 1 - T - R, \quad (3)$$

by measuring diffuse reflection (R) and diffuse transmission (T) using a commercial Lambda 950 spectrophotometer with integrating sphere from Perkin Elmer. The absorption of the nanodiamonds on a specific substrate was evaluated by subtracting the absorption of the twin samples with and without diamonds on the top [see TABLE I].

After determining all the components of Eq. (2), we can calculate the relative change in radiative decay rate of NV centers. For a gold substrate, there is suppression of emission $\Gamma/\Gamma_{ref} = 0.62$, while for the HMM, we observe enhancement $\Gamma/\Gamma_{ref} = 2.57$. From the dramatic relative changes of the measured total decay rates (inverse of the mean lifetimes) on gold substrate – 6.33 and on HMM – 13.48, it can be concluded that for these samples non-radiative decay rates are also substantially increased. A possible solution to this problem is to build HMMs using different designs[28] and low-loss constituent materials.[29]

In conclusion, NV centers in nanodiamonds are promising broadband single-photon sources capable of stable room temperature operation. In this work, we experimentally demonstrated the enhancement of spontaneous emission from NV centers over a broad range (broadband Purcell effect) by using an HMM consisting of alternating subwavelength-thick layers of gold and alumina. The achieved enhancement of the radiative decay rate on HMM substrate is about 3 times higher when compared to the reference sample with a bare glass substrate. In comparison to other proposed techniques for single-photon emission enhancement, our approach is based on a non-resonant way of engineering the electromagnetic environment which provides enhancement across the entire emission range of NV centers. In the future, we would like to achieve higher enhancement by building HMMs based on different designs and low-loss constituent materials. Such a diamond – metamaterial device can serve as a proof of principle for more complex structures that can bring quantum optical technologies to life.

The authors would like to thank N. Kinsey for his kind assistance with manuscript preparation, G. V. Naik for valuable discussions, and M. N. Slipchenko for the help with optical measurements. This work was partially supported by Air Force Office of Scientific Research grant FA9550-12-1-0024, U.S. Army Research Office grant 57981-PH (W911NF-11-1-0359) and NSF award Meta-PREM no. 1205457. SI is supported by the JSPS Postdoctoral Fellowships for Research Abroad.

[1] F. Jelezko and J. Wrachtrup, Phys. Status Solidi A **203**, 3207 (2006).

[2] C. Kurtsiefer, S. Mayer, P. Zarda and H. Weinfurter, Phys. Rev. Lett. **85**, 290 (2000).

[3] P. C. Maurer, G. Kucsko, C. Latta, L. Jiang, N. Y. Yao, S. D. Bennett, F. Pastawski, D. Hunger, N. Chisholm, M. Markham, D. J. Twitchen, J. I. Cirac and M. D. Lukin, Science **336**, 1283 (2012).

[4] F. Dolde, H. Fedder, M. Doherty, T. Nöbauer, F. Rempp, G. Balasubramanian, T. Wolf, F. Reinhard, L. Hollenberg, F. Jelezko and J. Wrachtrup, Nat. Phys. **7**, 459 (2011).

[5] J. R. Maze, P. L. Stanwix, J. S. Hodges, S. Hong, J. M. Taylor, P. Cappellaro, L. Jiang, M. V. Gurudev Dutt, E. Togan, A. S. Zibrov, A. Yacoby, R. L. Walsworth and M. D. Lukin, Nature **455**, 644 (2008).

[6] G. Balasubramanian, I. Y. Chan, R. Kolesov, M. Al-Hmoud, J. Tisler, C. Shin, C. Kim, A. Wojcik, P. R. Hemmer, A. Krueger, T. Hanke, A. Leitenstorfer, R. Bratschitsch, F. Jelezko and J. Wrachtrup, Nature **455**, 648 (2008).

[7] E. M. Purcell, Phys. Rev. **69**, 681 (1946).

[8] A. Faraon, P. E. Barclay, C. Santori, K.-M. C. Fu and R. G. Beausoleil, Nat. Photon. **5**, 301 (2011).

[9]S. Schietinger, T. Schroeder and O. Benson, Nano Lett. **8**, 3911 (2008).

[10]A. Faraon, C. Santori, Z. Huang, V. M. Acosta and R. G. Beausoleil, Phys. Rev. Lett. **109**, 033604 (2012).

[11]J. T. Choy, B. J. M. Hausmann, T. M. Babinec, I. Bulu, M. Khan, P. Maletinsky, A. Yacoby and M. Loncar, Nat. Photon. **5**, 738 (2011).

[12]H. N. S. Krishnamoorthy, Z. Jacob, E. Narimanov, I. Kretzschmar and V. M. Menon, Science **336**, 205 (2012).

[13]C. L. Cortes, W. Newman, S. Molesky and Z. Jacob, J. Opt. **14**, 063001 (2012).

[14]W. D. Newman, C. L. Cortes and Z. Jacob, J. Opt. Soc. Am. B **30**, 766 (2013).

[15]Z. Jacob, I. I. Smolyaninov and E. E. Narimanov, Appl. Phys. Lett. **100**, 181105 (2012).

[16]Z. Jacob, J.-Y. Kim, G. V. Naik, A. Boltasseva, E. E. Narimanov and V. M. Shalaev, Appl. Phys. B 100, 215 (2010).

[17]O. Kidwai, S. V. Zhukovsky and J. E. Sipe, Phys. Rev. A **85**, 053842 (2012).

[18]J. Kim, V. P. Drachev, Z. Jacob, G. V. Naik, A. Boltasseva, E. Narimanov and V. M. Shalaev, Opt. Express **20**, 8100 (2012).

[19]J. Elser, R. Wangberg, V. Podolskiy and E. Narimanov, Appl. Phys. Lett. **89**, 261102 (2006).

[20]M. A. Noginov, H. Li, Y. A. Barnakov, D. Dryden, G. Nataraj, G. Zhu, C. E. Bonner, M. Mayy, Z. Jacob and E. E. Narimanov, Opt. Lett. **35**, 1863 (2010).

[21]R. M. A. Azzam and N. M. Bashara, *Ellipsometry and polarized light* (North Holland, 1987).

[22]M. Y. Shalaginov, G. V. Naik, S. Ishii, M. N. Slipchenko, A. Boltasseva, J. X. Cheng, A. N. Smolyaninov, E. Kochman and V. M. Shalaev, Appl. Phys. B **105**, 191 (2011).

[23] Y.-R. Chang, H.-Y. Lee, K. Chen, C.-C. Chang, D.-S. Tsai, C.-C. Fu, T.-S. Lim, Y.-K. Tzeng, C.-Y. Fang, C.-C. Han, H.-C. Chang and W. Fann, Nat. Nanotechnol. **3**, 284 (2008).

[24] L.-H. Chen, T.-S. Lim and H.-C. Chang, J. Opt. Soc. Am. B **29**, 2309 (2012).

[25] M. N. Slipchenko, T. T. Le, H. Chen and J.-X. Cheng, J. Phys. Chem. B **113**, 7681 (2009).

[26] A. T. R. Williams, S. A. Winfield and J. N. Miller, Analyst (Lond.) **108**, 1067 (1983).

[27] W. Lukosz and R. E. Kunz, Opt. Commun. **20**, 195 (1977).

[28] A. V. Kildishev, A. Boltasseva and V. M. Shalaev, Science **339**, 1289 (2013).

[29] P. R. West, S. Ishii, G. V. Naik, N. K. Emani, V. M. Shalaev, and A. Boltasseva, Laser & Photon. Rev., 1 (2010).

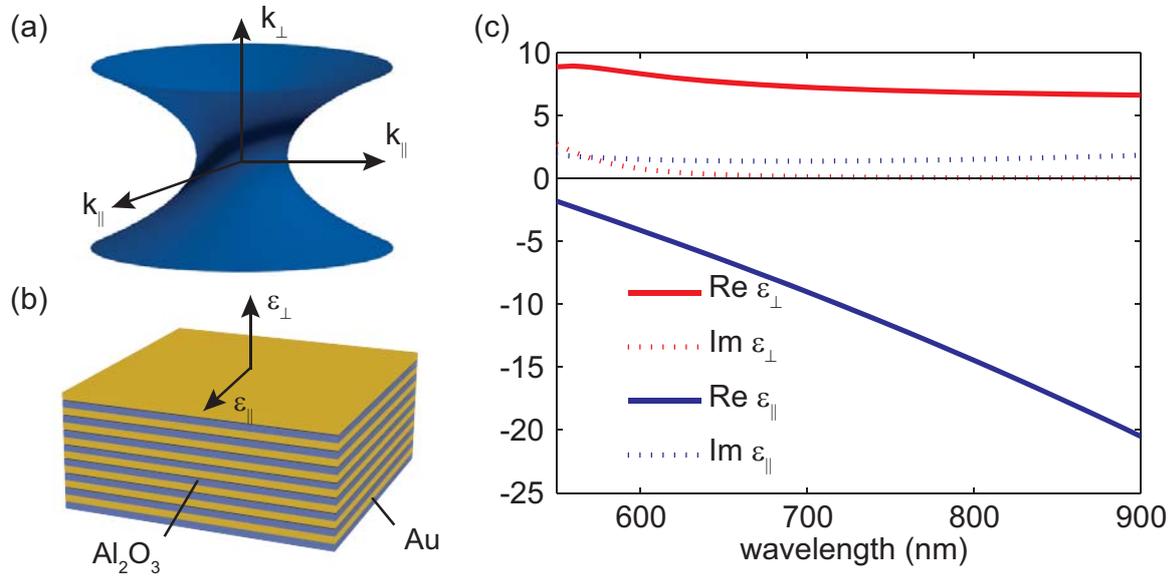

FIG. 1. Hyperbolic metamaterials (HMM). (a) Isofrequency surface of a hyperbolic metamaterial in k-space which supports waves with arbitrary large wavevector, ensuring extremely large photonic density of states. (b) Schematic of the fabricated HMM sample. The sample consists of 16 alternating gold and alumina layers, each 19 nm thick. The overall thickness is 304 nm. (c) Dielectric functions of fabricated HMM retrieved by spectroscopic ellipsometry measurements, within the range of the plot (550 – 900 nm) $\text{Re}\,\varepsilon_\perp > 0$, $\text{Re}\,\varepsilon_\parallel < 0$.

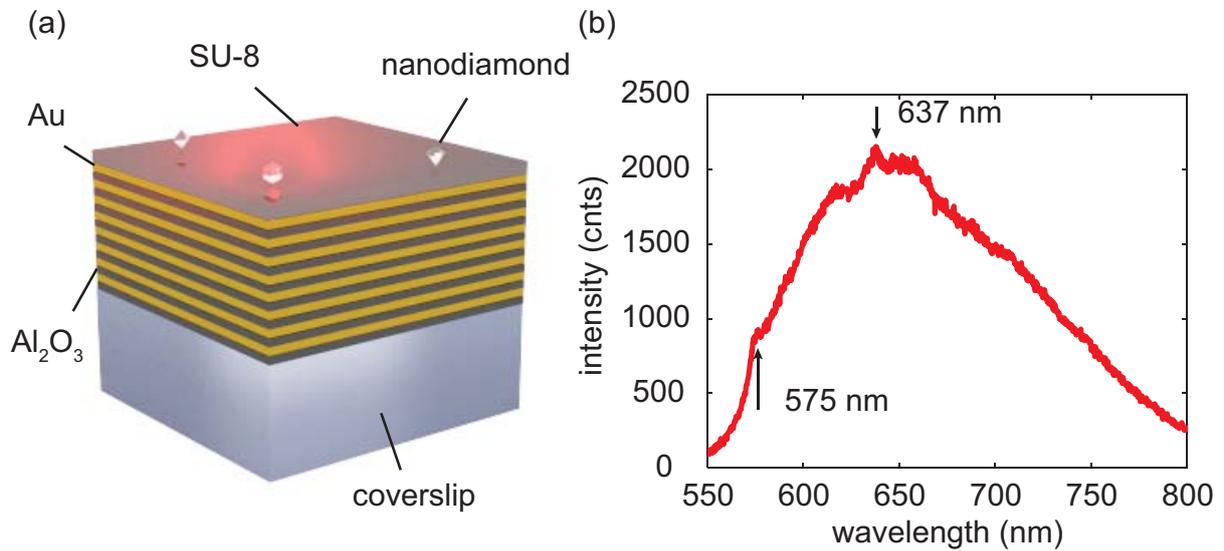

FIG. 2. Coupling of nitrogen-vacancy centers with HMM. (a) Schematic of the experimental sample consisting of HMM (16 alternating layers of Au and $Al_2O_3$, thickness of each layer - 19 nm), 30-nm-thick spacer layer composed of SU-8 photoresist, and nanodiamonds with NV centers inside. (b) Emission spectra of NV centers when nanodiamonds are placed on HMM. Excitation is 488-nm line of Ar/Kr continuous-wave laser (0.5 mW).

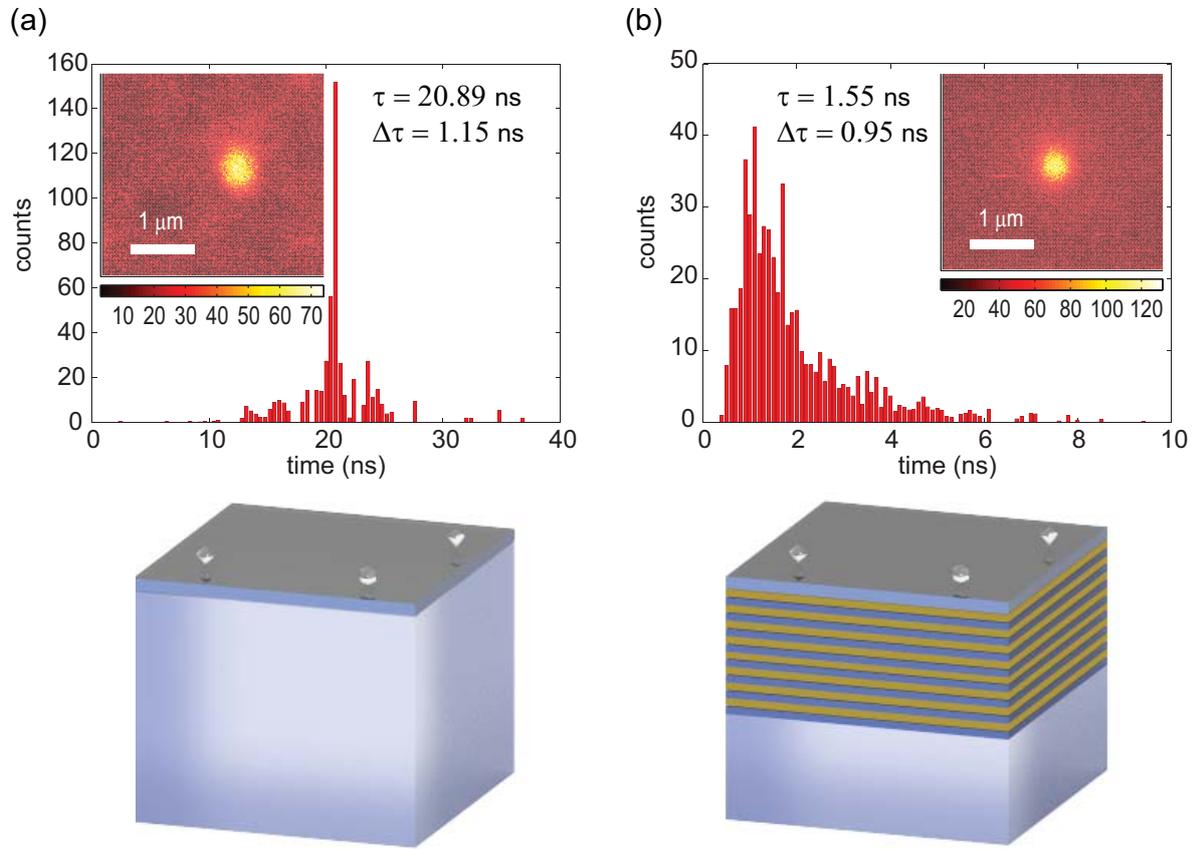

FIG. 3. Experimentally measured spontaneous emission lifetimes. Histograms of the NV center lifetimes on (a) coverslip and (b) HMM. Corresponding mean values (standard deviations) of the lifetime distributions are: (a) 20.89 ns (1.15 ns), (b) 1.55 ns (0.95 ns). HMM is deposited onto coverslip. In order to avoid photoluminescence quenching by metallic surface, spacer layer (diluted SU-8) is introduced between nanodiamonds and sample substrate.

TABLE I. Experimental results of measurements of lifetime (τ), integrated emission intensity (I), and absorption (A) for evaluation of relative radiative decay rates ($\Gamma/\Gamma_{ref}$) using Eq. (2).

| Sample substrate | τ, ns | $I/I_{ref}$ | A, % | $\Gamma/\Gamma_{ref}$ |
|---|---|---|---|---|
| HMM | 1.55 | 1.30 | 1.36 | 2.57 |
| Glass | 20.89 | 1.00 | 0.20 | 1.00 |